\documentclass[aps,prl,twocolumn,groupedaddress,showpacs]{revtex4}

\usepackage{graphicx}
\usepackage{amsmath}

\begin{document}

\title{Isotope Effect in the Superfluid Density of HTS Cuprates: Stripes, Pseudogap and
Impurities}

\author{J.L. Tallon$^1$$^,$$^3$, R.S. Islam$^2$, J. Storey$^3$, G.V.M. Williams$^1$ and J.R. Cooper$^2$}

\affiliation{$^1$MacDiarmid Institute, Industrial Research Ltd.,
P.O. Box 31310, Lower Hutt, New Zealand.}

\affiliation{$^2$IRC in Superconductivity, Cambridge University,
Cambridge CB3 0HE, England.}

\affiliation{$^3$School of Chemical and Physical Sciences,
Victoria University, Wellington, New Zealand.}

\date{\today}

\begin{abstract}

Underdoped cuprates exhibit a normal-state pseudogap, and their
spins and doped carriers tend to spatially separate into 1- or 2-D
stripes. Some view these as central to superconductivity, others
as peripheral and merely competing. Using
La$_{2-x}$Sr$_x$Cu$_{1-y}$Zn$_y$O$_4$ we show that an oxygen
isotope effect in $T_c$ and in the superfluid density can be used
to distinguish between the roles of stripes and pseudogap and also
to detect the presence of impurity scattering. We conclude that
stripes and pseudogap are distinct, and both compete and coexist
with superconductivity.
\end{abstract}


%%\begin{multicols}{2}
\maketitle

High-$T_c$ superconductors (HTS) remain a puzzle. Various
correlated states have been identified in HTS including
antiferromagnetism, the pseudogap\cite{Timusk}, nanoscale
spin-charge stripes\cite{Tranquada} and, of course,
superconductivity (SC). (Here we generalise``stripes" to include
possible 2D checkerboard structures\cite{Davis}). The pseudogap is
a nodal energy gap of uncertain origin that appears in the
normal-state (NS) density of states (DOS). Its effects can be
observed in many physical
properties\cite{Timusk,Talloram}. %Stripes originate from the
%spatial separation of doped holes from the background Cu spins
%resulting in strips of antiferromagnetic domains separated by
%linear domain walls on which the charge resides. They are observed
%primarily by inelastic neutron scattering\cite{Tranquada} or
%NMR\cite{Singer}.
Several opposing views are still current. One is that stripes play
a central role\cite{Emery1}, forming the pseudogap
correlation\cite{Salkola} and/or mediating the SC pairing. Another
is that the NS pseudogap arises from incoherent superconducting
fluctuations which set in well above $T_c$\cite{Emery2}. Another
is that these states are independently competing\cite{Loram}. Here
stripes and pseudogap play a secondary role and SC is mediated by
some other pairing boson. An unambiguous test of these opposing
views is urgently needed. We show here that isotope effects
provide such a test.

The isotope exponent $\alpha(E)$ in a given property $E$ is
defined as $\alpha(E)$ = $-$ ($\Delta E/E$)/($\Delta M/M$), where
$M$ is the isotopic mass and $E$ may be $T_c$, the SC gap
parameter, $\Delta_0$, the pseudogap energy scale, $E_g$, or the
superfluid density $\rho_s$ = $\lambda_{ab}^{-2}$ = $\mu _0 e^2
(n_s/m_{ab}^*)$. ($\lambda_{ab}$ is the in-plane London
penetration depth, $n_s$ is the carrier density and $m_{ab}^*$ is
the effective electronic mass for in-plane transport).
%The negative sign reflects the fact that $E$ is usually an energy
%scale which is reduced by an increase in isotopic mass.
An isotope effect on $T_c$ was
%contemplated, though not observed, by Kammerlingh Onnes\cite{KO}.
%Thirty years were to pass before its eventual discovery in Sn by
first discovered in 1950 by Allen {\it et al}. for Sn\cite{Allen}.
They found $\alpha(T_c) \approx 0.5 \pm 0.05$ which provided the
central clue for the role of phonons in pairing and led 7 years
later to the BCS theory of SC\cite{BCS}.

The situation with HTS is more complex. The oxygen isotope effect
on $T_c$ was found\cite{Batlogg} to be small, with $\alpha(T_c)
\approx 0.06$. However, with decreasing doping the effect rises
and eventually diverges as $T_c \rightarrow
0$\cite{Franck,Pringle}. Surprisingly, an isotope effect was also
found in the superfluid density\cite{Zhao} (and attempts were made
to resolve this into a dominant isotope effect just in
$m^*$\cite{Zhao2,Zhao1}). We will show that both of these unusual
effects can be understood in terms of a normal-state pseudogap
which competes with SC\cite{Williams}. We also predict and confirm
an isotope effect in $\rho_s$ induced by impurity scattering. The
isotope effects in $T_c$ and $\rho_s$ are mapped as a function of
doping in La$_{2-x}$Sr$_x$Cu$_{1-y}$Zn$_y$O$_4$ and we observe a
canonical pseudogap behavior as well as a huge anomalous effect
associated with stripes. The clear distinction between these
effects shows that the pseudogap and stripe states are distinct
and both compete with SC.

An isotope effect, $\alpha(\rho_s)$, in the superfluid density is
surprising because for a simple BCS superconductor it is
rigorously zero. According to Leggett's theorem, $\rho_s$ is just
the total integrated spectral weight of the free carriers i.e. the
total carrier density divided by the effective mass\cite{Leggett}.
But, when there are strong departures from nearly-free-electron
theory this need not be so. We identify two cases for HTS in which
an isotope effect in $\rho_s$ arises: in the presence of (i)
impurity scattering, and (ii) a pseudogap.

HTS possess a $d$-wave order parameter and in the presence of {\it
impurity scattering} both $T_c$ and $\rho_s$ are diminished. The
degree to which they are reduced depends upon the magnitude of the
scattering rate, $\Gamma$, relative to the maximum gap parameter,
$\Delta_0$, near ${\bf k}=(\pi,0)$. In the presence of a competing
{\it pseudogap}, spectral weight removed by the pseudogap is no
longer available for the condensate and, again, both $T_c$ and
$\rho_s$ are diminished. The degree to which they are reduced
depends upon the relative magnitudes of the pseudogap and the SC
gap.
%The former is shown from specific heat,
%NMR, ARPES, tunneling and thermal conductivity to be a nodal gap
%which grows with underdoping\cite{Timusk,Talloram}. A simple
%triangular gap is a good approximation.
Thus, $T_c$ and $\rho_s$ are reduced according to the magnitude of
the ratios $\Gamma/\Delta_0$ for impurity scattering, and
$E_g/\Delta_0$ for a pseudogap. In either case, a relatively small
isotope effect in $\Delta_0$ will necessarily produce enhanced
isotope effects in $T_c$ and $\rho_s$ which diverge as $T_c
\rightarrow 0$. Now it has been shown from specific heat, NMR and
ARPES that, with increasing doping, $E_g$ decreases and closes
abruptly at a critical doping state, $p_{crit}=0.19$ holes/Cu, in
the lightly overdoped regime\cite{Talloram}. It follows that the
isotope effect in $\rho_s$ should disappear at critical doping
where the pseudogap closes provided that impurity scattering is
absent. Our initial task is to quantify these effects.

First, we recall that there is no isotope effect in the pseudogap.
We have examined the $^{89}$Y Knight shift in YBa$_2$Cu$_4$O$_8$
using magic angle spinning with extremely narrow linewidths
($\approx$100 Hz) and found no isotope effect within the bounds
$\alpha(E_g) \leq 0.01$\cite{Williams}. Though not essential, we
proceed under the assumption that an isotope effect is confined to
the pairing gap, $\Delta_0$, and absent from the pseudogap, $E_g$.
The small isotope effect observed in $1/T_1T$ in the same compound
does not reflect an isotope effect in the pseudogap. Using the
enhanced susceptibility formalism it devolves, rather
surprisingly, into an isotope effect in the paramagnon
frequency\cite{Talloram}.

\begin{figure}
\centerline{\includegraphics*[width=70mm]{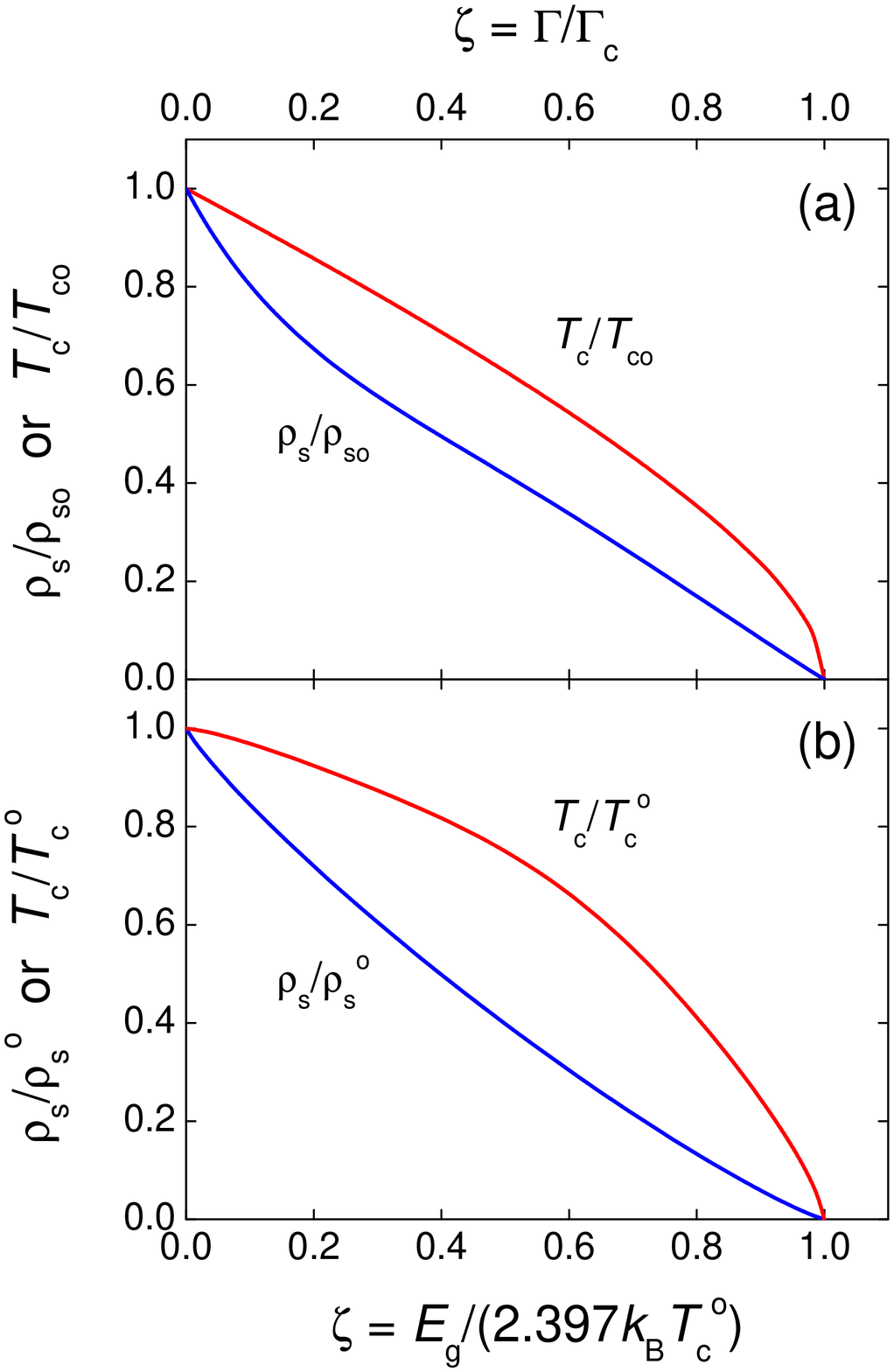}}
\caption{\small The fractional suppression of $T_c$ and superfluid
density in a $d$-wave SC with (a) unitary-limit impurity
scattering and (b) a NS triangular pseudogap with gap energy
$E_g$.}
\end{figure}

{\it Theory}. Impurity scattering for a $d$-wave order parameter
has been investigated by many authors\cite{Maki,Tallon}. We
summarize the results in Fig. 1(a) which shows the depression of
$T_c$ and $\rho_s$ as a function of $\zeta=\Gamma/\Gamma_c$, where
$\Gamma_c$ is the critical scattering rate for fully suppressing
$T_c$. The reduction in $T_c$ follows the standard
Abrikosov-Gorkov equation. In the unitary limit, the scattering
rate $\Gamma= n_i/\pi N(E_F)$ where $n_i$ is the density of
scatterers and $N(E)$ is the NS DOS. %The rate of suppression
%increases abruptly with the opening of the pseudogap due to the
%associated fall in NS DOS\cite{Tallon}.
Fig. 1(a) shows that $T_c(\zeta)/T_{c0}$ falls at first slowly
then accelerates while $\rho_s(\zeta)$ falls at first rapidly then
slows as $\Gamma$ grows. We define the functions $h$ and $g$ given
by $\rho_s(\zeta)/\rho_{s0} = h(\zeta)$ and $T_c(\zeta)/T_{c0} =
g(\zeta)$. The isotope effects in $\rho_s$ and $T_c$ are
\begin{eqnarray}
\nonumber \alpha(\rho_s) & = & - \zeta (h^{\prime}/h)
\alpha(T_{c0})
\\ \alpha(T_c) & = & {\bf [}1 - \zeta (g^{\prime}/g){\bf ]}
\alpha(T_{c0})
\end{eqnarray}

\noindent The prime indicates differentiation of $h(\zeta)$ or
$g(\zeta)$. Thus
\begin{equation}
\ \alpha(\rho_s) = - \zeta (h^{\prime}/h) {\bf [}1 - \zeta
(g^{\prime}/g){\bf ]}^{-1} \alpha(T_{c})
\end{equation}

We have shown previously\cite{Pringle,Williams} that the isotope
effect in $T_c$ across the entire phase diagram is consistent with
an underlying exponent (in the absence of scattering and
pseudogap) of $\alpha(T_{c0}) \approx 0.06$. The red line in Fig.
2(a) shows $\alpha(\rho_s)$ plotted versus $\alpha(T_c)$ using
this value. In the absence of impurity scattering $\alpha(\rho_s)$
= 0 and $\alpha(T_c)$ = $\alpha(T_{c0})$ = 0.06. This is the
left-hand termination of the red line. With increasing scattering
both $\alpha(\rho_s)$ and $\alpha(T_c)$ rise along the
line, and finally diverge as $\Gamma \rightarrow \Gamma_c$. %We
%will see that this is satisfied experimentally.

\begin{figure}
\centerline{\includegraphics*[width=80mm]{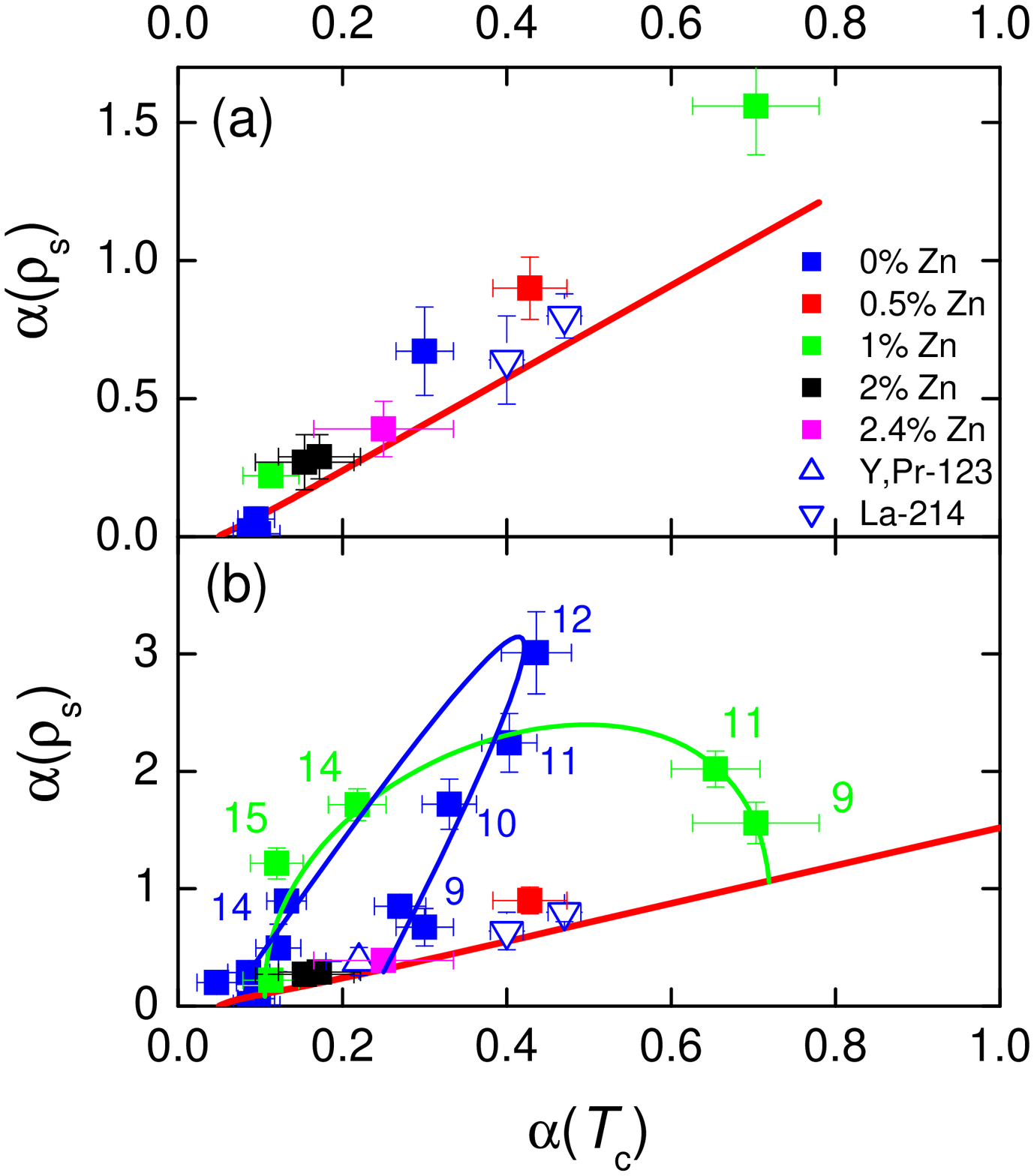}}
\caption{\small Oxygen isotope exponent in superfluid density
plotted against that in $T_c$ for
La$_{2-x}$Sr$_x$Cu$_{1-y}$Zn$_y$O$_4$. Panel (a) shows the
calculated (red line) and observed effect of impurities for
overdoped ($x=0.19$, $y=0$, 1, 2, 2.4\%) and strongly underdoped
($x=0.09$, $y=0$, 0.5, 1\%) samples.  Previous data is also shown
(blue down triangles). Panel (b) shows more data for $0.10 \leq x
\leq 0.19$, with $y=0$ (blue squares), and 1\% Zn (green squares)
and reveals an anomalous deviation from the canonical pseudogap
line (red line). Previous data is shown for
Y$_{1-z}$Pr$_z$Ba$_2$Cu$_3$O$_{7-\delta}$ (blue up triangle).}
\end{figure}

Turning to the pseudogap, specific heat\cite{Loram} and tunneling
measurements\cite{Krasnov} show that the pseudogap is
non-states-conserving, with an approximately triangular energy
dependence, and pinned to the Fermi level, $E_F$. We assume
therefore a triangular normal-state DOS:
\begin{equation}
\begin{array}{rclr}
N(E) & = & N_0 \times |E - E_F|/E_g(p) ; & |E - E_F| < E_g(p), \\
& = & N_0 ; & |E - E_F| > E_g(p).
\end{array}
\end{equation}

\noindent and solve standard weak-coupling $d$-wave BCS
expressions to calculate $T_c$ as a function of $E_g$. For this
particular NS DOS $T_c \rightarrow 0$ as $E_g \rightarrow
2.397k_BT_c^0$ where $k_B$ is Boltzmann's constant and $T_c^0 =
T_c(E_g=0)$. Fig. 1(b) shows $T_c$ plotted as a function of $\zeta
= E_g/(2.397k_BT_c^0)$. As for impurity scattering, the depression
in $T_c$ is slow at first and more rapid as $\zeta  \rightarrow
1$.

Elsewhere\cite{Tallon2} we have calculated the effect of a
triangular pseudogap on $\rho_s$. The approach is admittedly for a
Fermi liquid but we note that the effects we describe are
dominated by the nodal regions of the Fermi surface where such a
Fermi liquid approach is more likely to be valid. $\rho_s(\zeta)$
is plotted as a function of $\zeta$ in Fig. 1(b). This exhibits an
initial rapid fall which slows as $E_g$ grows and $\zeta
\rightarrow 1$. The overall behavior is qualitatively similar to
that shown in Fig. 1(a) for impurity scattering, but differs in
detail. We could therefore define new functions $h(\zeta)$ and
$g(\zeta)$ as above and derive an equation formally identical to
eq. (2) to describe the isotope effects in $\rho_s$ and $T_c$
associated with the presence of the pseudogap. These equations
show that when the pseudogap closes at critical doping we have
$\zeta=0$ and $\alpha(\rho_s) = 0$ while $\alpha(T_c) =
\alpha(T_c^0)$. The resultant curve $\alpha(\rho_s)$ versus
$\alpha(T_c)$ almost exactly coincides with the red line in Fig.
2(a). If there were an isotope effect in the pseudogap
$\alpha(T_{c0})$ in eq. (1) should be replaced by
$[\alpha(T_{c0})- \alpha(E_g)]$ and eq. (2) and the red line in
Fig. 2 remain unchanged.

{\it Experimental details.} La$_{2-x}$Sr$_x$Cu$_{1-y}$Zn$_y$O$_4$
samples were synthesized by solid state reaction at 985$^\circ$C
in air by repeated milling, pelletization and reaction until phase
pure as determined by x-ray diffraction. Two small, approximately
$2 \times 2 \times 3$ mm$^3$ bars, were cut from alongside each
other at the centre of each of the resultant pellets to ensure, as
much as possible, identical pairs. They were isotope exchanged in
identical quartz tubes, one charged with $^{16}$O and the other
with $^{18}$O, side by side in a furnace. The $^{18}$O gas (from
Isotec) was 99\% enriched and several exchanges were employed
until about 95\% exchange was achieved. On the final exchange the
samples were slow cooled then annealed for 15 hours at
500$^\circ$C to ensure oxygenation to full stoichiometry. The
degree of exchange was confirmed by Raman measurements of the
spectral shifts of the oxygen phonons.

To determine the isotope shifts in $T_c$ and $\rho_s$ we carried
out field-cooled DC magnetization measurements in the mixed state
at 150 Oe. For this regime Zhao and Morris\cite{Zhao2} adopted the
relation\cite{Finnemore}
\begin{equation}
\ (-M)^{1/2}  \propto  (r_g/\lambda) {\bf [}T_c(H)-T{\bf ]} \times
{\bf [} |dH_{c2}/dT| /(2\kappa^2-1) T_c{\bf ]}^{1/2}
\end{equation}

\noindent for the limit, near $T_c$, of $\lambda \gg r_g$. Here
$r_g$ is the mean radius of the SC grains. These authors showed
that this relation could be used to deduce separate isotope
effects in $n_s$ and $m^*$. But the algebra was incorrect (see
Appendix below). A further problem arises\cite{Clem} in that, for
small particles, this relation does not satisfy the sum rule,
$\mu_0 \int M(H)dH = U_0 \equiv$ the condensation energy. With a
mean grain size of 25$\mu$m and $\lambda_{ab}(0)$ ranging from 0.2
to 0.32$\mu$m\cite{Panag}, we adopt the limit $\lambda \ll r_g$
which is clearly satisfied up to a few K below $T_c$. This yields
\begin{equation}
\ -M  \propto  \lambda^{-2} {\bf [}1 - T/T_c{\bf ]} .
\end{equation}

\noindent Thus the isotope coefficient in the slope of $-M(T)$ is
given by $\alpha(\rho_s)-\alpha(T_c)$. What we report is the
partial isotope exponent due to the change in oxygen mass only. We
obtained qualitatively similar results with Meissner state
measurements at 10 Oe (not shown).

{\it Results}. We start first with the effect of impurity
scattering in the overdoped region $x=0.19$ where the pseudogap is
absent. Illustrative plots of magnetization versus temperature are
shown in Fig. 3(a) for $y$ = 0, 1, 2 and 2.4\% (solid curves). It
evident that an isotope effect in $T_c$ is present in each but
that one in $\rho_s$ is only present in the higher Zn
concentrations. Values of $\alpha(\rho_s)$ are plotted versus
$\alpha(T_c)$ in Fig. 2 (a) (left-hand cluster of blue, green,
black and mauve squares) and they are seen to be roughly
consistent with the model calculation. The fact that
$\alpha(\rho_s) \rightarrow 0$ as $y \rightarrow 0$ indirectly
shows that any disorder potential present in Zn-free
La$_{2-x}$Sr$_x$CuO$_4$ in the overdoped region is too small to
present significant scattering and hence $\alpha(\rho_s) = 0$. It
also seems unlikely that there is any significant phase separation
because the domain walls would surely act as scattering centers.

\begin{figure}
\centerline{\includegraphics*[width=80mm]{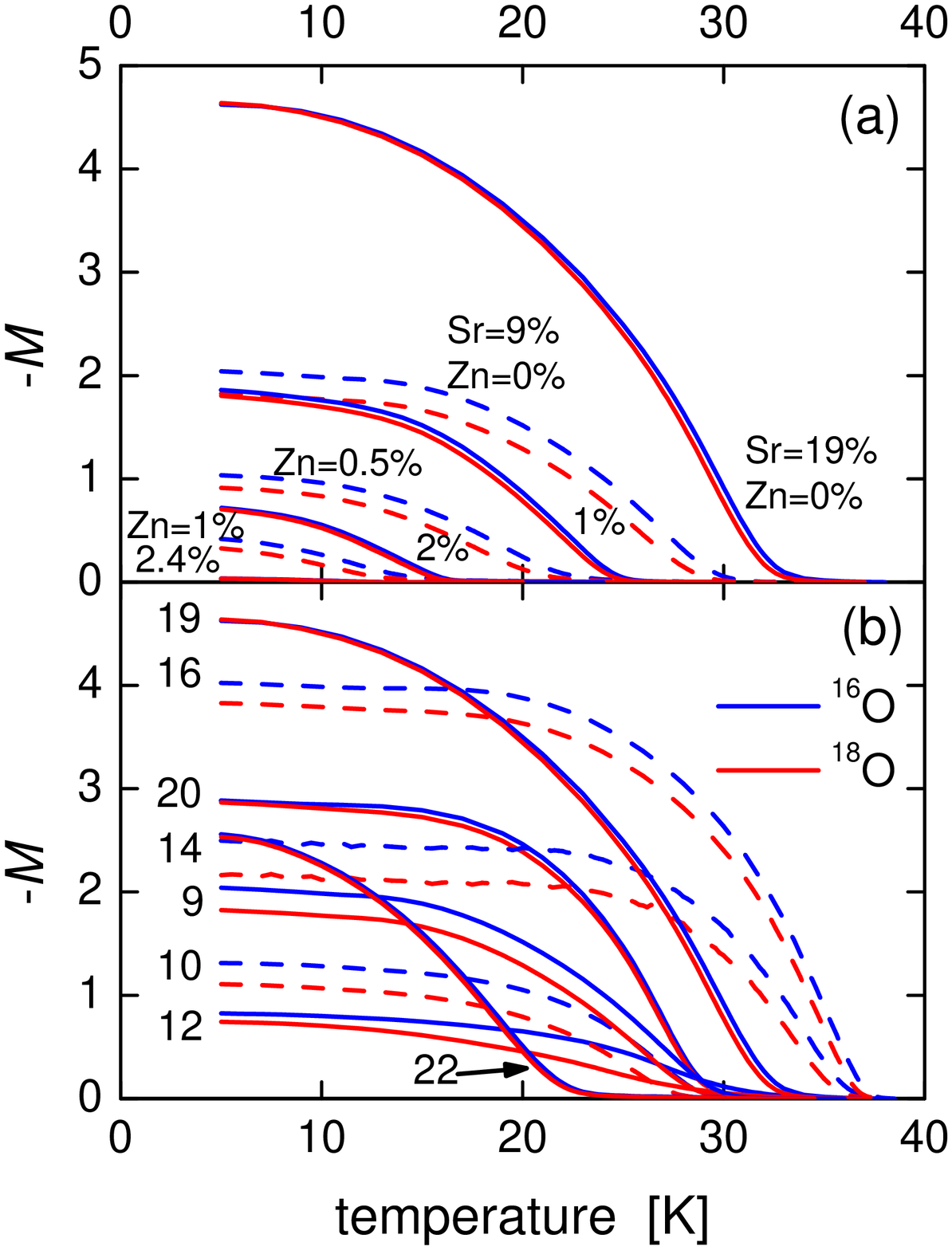}}
\caption{\small Magnetization versus $T$ for
La$_{2-x}$Sr$_x$Cu$_{1-y}$Zn$_y$O$_4$. Panel (a) shows $x=0.19$
with $y=0$, 1, 2, 2.4\% (solid curves) and $x=0.09$ with $y=0$,
0.5, 1\% (dashed curves).  Panel (b) shows data for Zn-free
samples with $0.09 \leq x \leq 0.22$.}
\end{figure}

Turning to the heavily underdoped pseudogap region, we show
magnetization curves in Fig. 3(a) for $x=0.09$ with $y=0$, 0.5 and
1\% Zn (dashed curves). The resultant $\alpha(\rho_s)$ versus
$\alpha(T_c)$ values are plotted in Fig. 2(a), shown by the blue,
red and green data points to the right. These continue to track up
the canonical curve, showing that the pseudogap and impurity
scattering have essentially the same effect in such a plot.
%, just as we have calculated.
To this data we add previously-reported\cite{Hofer} values for
La$_{2-x}$Sr$_x$CuO$_4$ obtained using muon spin relaxation
($\mu$SR) with $x$ = 0.080 and 0.086 (blue down triangles). The
collective data is generally consistent with the model.
% but noticeably lies a little to the left of the theoretical line. The
%reason for this becomes obvious when we explore the intermediate
%doping region in the neighborhood of 1/8 doping.

Fig. 3(b) shows a selection of illustrative plots of $M$ vs $T$
for Zn-free samples with $x$ ranging from 0.09 to 0.22. It is
immediately evident from the low-$T$ values of $M$ that
$\alpha(\rho_s)$ = 0 for all $x > 0.19$ but becomes non-zero and
large as $x$ falls below 0.19. Values of $\alpha(\rho_s)$ are
plotted against $\alpha(T_c)$ in Fig. 2 (b) (blue squares) and
increasing doping is shown by the arrow. Here, a remarkable
anomaly is evident. The overdoped data and the heavily underdoped
data lie near the canonical pseudogap line. But near $x=0.12$ the
data deviates drastically from this canonical behavior. This is
presumably due to the presence of charged stripes, inferred from
neutron scattering near $p$=1/8, which provide strong electronic
coupling to the lattice. If the pseudogap itself arose from
fluctuating stripes one might expect the anomaly to drive up the
canonical line. The huge deviation suggests a fundamentally
different behavior and clearly distinguishes stripes from the
pseudogap near $p$=1/8.

In order to further test this interpretation we examined the
effects of non-magnetic Zn substitution. Our expectation was that
the combined effects of spin vacancies and the tendency of Zn to
enhance the canonical behavior would be to broaden and weaken the
anomaly pushing it up the canonical line. Fig. 2(b) shows the
effect of 1\% Zn substitution (green squares). The contour,
indicated by the green curve, confirms our expectations.

Finally, we show by the upward open triangle in Fig. 2(b) recently
reported $\alpha(\rho_s)$ and $\alpha(T_c)$ data for
Y$_{1-z}$Pr$_z$Ba$_2$Cu$_3$O$_{7-\delta}$ obtained by Khasanov
{\it et al}. using $\mu$SR\cite{Khasanov}. For $z = 0.3$ these
authors found $T_c = 59.3$K and we estimate that the doping state
is very close to $p = 0.125$. And yet the data resides close to
the canonical line completely free of the anomalous deviation
associated with stripes. It is clear from inelastic neutron
scattering studies that the YBa$_2$Cu$_3$O$_{7-\delta}$ compound
exhibits a much weaker tendency to stripe formation. Consistent
with this we find this sample exhibits essentially stripe-free
canonical pseudogap behavior.

We conclude that our results and analysis demonstrate a clear
distinction between the canonical effects on the superfluid
density arising from the pseudogap and impurity scattering on the
one hand and stripe correlations on the other. We achieve this by
examining a plot of $\alpha(\rho_s)$ versus $\alpha(T_c)$ which is
relatively insensitive to the precise details of the NS DOS.
Stripes cause a huge deviation from this canonical behavior
associated with the strong electronic coupling to the lattice
arising from spatial charge modulation. On the basis of these
results we make the important conclusion that stripe and pseudogap
correlations are fundamentally different and both compete with
each other and with superconductivity. %This also means that the
%pseudogap is unrelated to superconducting fluctuations. The
%absence of an isotope effect in superfluid density at and above
%critical doping suggests that Zn-free La$_{2-x}$Sr$_x$CuO$_4$ is
%essentially free of disorder potential scattering in this region
%and we may further infer the absence of nanoscale phase
%separation. These important conclusions highlight the powerful
%insights obtainable from isotope effect studies.

We acknowledge financial support from the Marsden Fund and the
MacDiarmid Institute (JLT, JS and GVMW) and from Trinity College,
Cambridge and the Cambridge Commonwealth Trust (RSI).

\section*{Appendix - isotope effect in {\it\symbol{'155}}$^*$?}
Several authors\cite{Zhao1,Zhao2} have considered the possibility
that the isotope effect in $\rho_s \propto n_s/m^*$ may be
resolved into $\alpha(n_s)$ - $\alpha(m^*)$ and they have sought
to determine these two components separately.

Zhao {\em et al.}\cite{Zhao1} investigated the oxygen isotope
dependence of the orthorhombic/tetragonal (O/T) transition in
La$_{2-x}$Sr$_x$CuO$_4$ and found a null effect. Because the O/T
transition temperature is doping dependent they took this to
indicate that there was no isotope effect in the carrier
concentration and consequently the isotope effect in $\rho_s$
derives wholly from the isotope effect in $m^*$ i.e.
$\alpha(\rho_s)$ = $-\alpha(m^*)$. However, the location of the
O/T transition is an ion-size dependent effect not primarily a
doping effect and, moreover, there is no simple relationship
between the doped hole concentration, $x$, and the carrier
concentration.

Elsewhere, Zhao and Morris\cite{Zhao2} use eq. (5) for 10G
measurements to yield a magnetisation slope
\begin{equation}
\ P_1  \propto  r_g^2\, n_s/T_c\, m^* ,
\end{equation}
while for 150G measurements they use eq. (4) from which they
deduce
\begin{equation}
\ P_2  \propto  r_g\, n_s^{5/3}/T_c\, m^* .
\end{equation}
Clearly, measurement of the isotope effects in $P_1$ and $P_2$
would allow extraction of the individual isotope effects in $n_s$
and in $m^*$. However, eq. (4) does not lead to eq. (7).

To see this we consider the relation\cite{Werthamer}
\begin{equation}
\ H_{c2}(0) = 0.7\, T_c\, [dH_{c2}/dT]_{T_c} = \phi_0/(2\pi
\xi(0)^2) .
\end{equation}
which, on substitution in eq.(4) when $\kappa=\lambda(0)/\xi(0)
\gg 1$, reduces to eq. (5) and
\begin{equation}
\ P_2  \propto P_1 \propto n_s/T_c\, m^* .
\end{equation}
Thus the isotope effect in $\rho_s$ cannot be separated into
separate contributions from $\alpha(n_s)$ and $\alpha(m^*)$ in the
way suggested by Zhao and Morris.

\end{document}